\documentclass[iop, apjl]{emulateapj}
\usepackage[backref,breaklinks,colorlinks,citecolor=blue]{hyperref}
\usepackage{amsmath,amstext}
\usepackage[T1]{fontenc}
\usepackage{apjfonts}
\usepackage[figure,figure*]{hypcap}
 
\usepackage{graphicx}
\usepackage{color}
\DeclareGraphicsExtensions{.eps,.png,.jpg}
\usepackage{epsfig, psfig}
\usepackage[normalem]{ulem}
\usepackage{soul}

\def\amin{$^{\prime}$}
\def\asec{$^{\prime\prime}$}

\def\hst{{\it HST}}
\def\kms{km s$^{-1}$}

\def\lax{{$\mathrel{\hbox{\rlap{\hbox{\lower4pt\hbox{$\sim$}}}\hbox{$<$}}}$}}
\def\gax{{$\mathrel{\hbox{\rlap{\hbox{\lower4pt\hbox{$\sim$}}}\hbox{$>$}}}$}}
\def\simlt{\lower.5ex\hbox{$\; \buildrel < \over \sim \;$}}
\def\simgt{\lower.5ex\hbox{$\; \buildrel > \over \sim \;$}}

\def\sb{mag~arcsec$^{-2}$}

\shortauthors{Gu et al.} 
\begin{document}
\title{Coordinated Assembly of Brightest Cluster Galaxies}


\author{
  Meng Gu\altaffilmark{1}, 
  Charlie Conroy\altaffilmark{1} 
  and Gabriel Brammer\altaffilmark{2}
   }                         
\altaffiltext{1}{Department of Astronomy, Harvard University, Cambridge, MA 02138, USA}
\altaffiltext{2}{Space Telescope Science Institute, 3700 San Martin Drive, Baltimore, MD 21218, USA}
\submitted{Accepted for publication in ApJL}
\begin{abstract}

Brightest Cluster Galaxies (BCGs) in massive dark matter halos are shaped 
by complex merging processes. We present a detailed stellar population 
analysis in the central region of Abell~3827 at $z\sim0.1$, including 
five-nucleus galaxies involved in a BCG assembly.  Based on deep 
spectroscopy from Multi Unit Spectroscopic Explorer (MUSE), 
we fit the optical spectra of 13 early-type galaxies (ETGs) in the central 
$70$~kpc of the cluster.  The stellar populations in the central $R=1$~kpc of 
these ETGs are old ($>6$~Gyr).  Their [Fe/H] increases with 
$\sigma_{\star}$ and stellar mass.  More importantly, [$\alpha$/Fe] of 
galaxies close to the cluster center do not seem to depend on 
$\sigma_{\star}$ or stellar mass, indicating that the cluster center 
shapes the [$\alpha$/Fe]--$\sigma_{\star}$ and [$\alpha$/Fe]--$M_{\star}$ 
relations differently than other environments where [$\alpha$/Fe] is 
observed to increase with increasing $\sigma_{\star}$ or $M_{\star}$. Our 
results reveal the coordinated assembly of BCGs: their building blocks 
are different from the general low mass populations by their high 
[$\alpha$/Fe] and old ages.  Massive galaxies thus grow by accreting preferentially 
high [$\alpha$/Fe] and old stellar systems.  The radial profiles also bear the imprint 
of the coordinated assembly.  Their declining [Fe/H] and flat [$\alpha$/Fe] 
radial profiles confirm that the accreted systems have low metallicity 
and high [$\alpha$/Fe] stellar contents.   
\end{abstract}

\maketitle
 \section{Introduction}

According to the $\Lambda$--Cold Dark Matter model, galaxy assembly  
is closely linked to the hierarchical growth of dark matter structures.  
Local massive early-type galaxies (ETGs) are considered to have 
evolved from the compact ``red nuggets'' at $z\approx2$ by doubling their 
stellar masses and increasing their effective radii by a factor of $3-5$ 
\citep[e.g.][]{vanDokkum2010, Patel2013}.  Recent simulations describe this 
transformation by the two-phase scenario \citep[e.g.][]{Naab2009, Oser2010, 
Oser2012}, in which massive ETGs experience strong dissipational 
processes that lead to rapid and concentrated mass growth at high redshifts, 
and accrete low mass systems at later times to build up the outer envelopes. 
Brightest cluster galaxies (BCGs) are a special class of ETGs at the extreme 
high-mass end of the stellar mass function and in the densest environments.  
They have diffuse and extended envelops 
\citep[e.g.][]{Schombert1988} that can be explained by a series of merging 
events \citep[e.g.,][]{Ostriker1975, Hausman1978, Dubinski1998}.  

It is still ambiguous whether the low mass galaxies we observe today are 
intrinsically different from the building blocks of massive galaxies, 
or their surviving counterparts.  One useful approach is to compare the 
abundance trends of their stellar contents, especially the ratio between 
$\alpha-$elements to iron.  [$\alpha$/Fe] is used to indicate the star 
formation timescales due to its sensitivity to the time delay between 
SNe~II and SNe~Ia \citep[e.g.][]{Tinsley1979}.  SNe~II from massive stars 
yield both $\alpha$--elements and Fe, while SNe~Ia from low mass binary 
systems contribute mostly Fe on longer timescales.  High [$\alpha$/Fe] in old 
stellar population suggests short star formation timescales in the past.
Stellar population analysis has revealed correlations between stellar 
population properties and stellar mass or stellar velocity dispersion 
\citep[e.g.][]{Trager2000, Worthey2003, Thomas2005, Schiavon2007, 
Thomas2010, Conroy2014, McDermid2015} 
in a way that massive galaxies are older, 
more metal rich and more $\alpha$--enhanced compared to low mass galaxies. If 
these trends are universal for all environments and epochs, an apparent 
tension under the hierarchical assembly paradigm would emerge: the building 
blocks of massive galaxies, especially the BCGs would have ``diluted'' the 
[$\alpha$/Fe] at the high mass end, and/or would produce steep 
[$\alpha$/Fe] gradients. This would make it difficult to reconcile with the general 
observational facts that more massive galaxies are more $\alpha$--enhanced.  

\vskip 0.1cm 
\begin{figure*}
  \centerline{\psfig{file=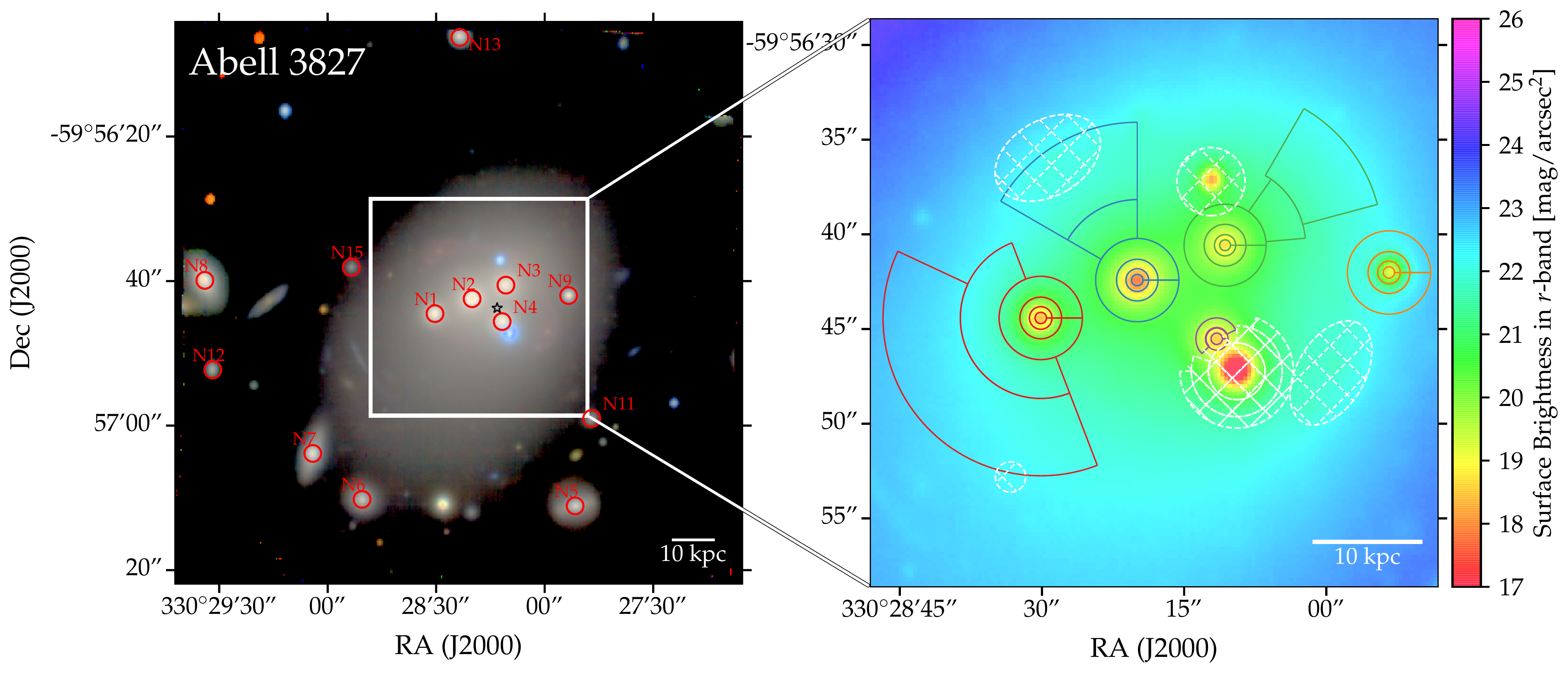,width=18.0cm}}
  \figcaption[A38272015_Figure1_May9.pdf]{ 
    Left panel: Overview of the locations of ETGs studied in this paper on a 
    SDSS--$riz$ composite image derived from the MUSE datacube.  Red circles 
    enclose $R=2$~kpc.  
    Right panel:  Zoomed-in SDSS--$r$ band surface brightness map derived from 
    the MUSE datacube.  Solid lines enclose stacked regions for radial profiles.  White hatched regions highlight masked out spaxels.
    }
\end{figure*}
\noindent 

The Milky Way satellite dwarf galaxies were once thought to be the surviving 
counterparts of Galactic building blocks until studies revealed their stellar 
populations occupy different locations from Milky Way halo stars on the 
[$\alpha$/Fe] vs. [Fe/H] diagram \citep[e.g.][]{Tolstoy2009}. Looking beyond 
the Milky Way, \citet{Liu2016} showed that the [$\alpha$/Fe] of 
low mass ETGs in the Virgo cluster depend on the distance to the cluster center 
and the [$\alpha$/Fe]--$\sigma_{\star}$ relation in this cluster has larger 
scatter, indicating that the densest environments quench low mass galaxies earlier 
than other environments.  From these studies it seems that the assembly of massive 
galaxies are coordinated in a way that their building blocks have early truncated 
star formation histories, making them a particular sample with high [$\alpha$/Fe] 
among the low mass systems.  

In this Letter, we present stellar population analysis on 13 ETGs in 
Abell~3827.  Five of them are involved in a rarely observed BCG assembly from 
multiple mergers, therefore we are fortunate to directly analyze the 
building blocks of a prospective BCG.  [Mg/Fe] is used as a tracer of [$\alpha$/Fe]. 
We compare the stellar population scaling 
relations in this special environment to the general samples in 
previous work.  We assume a flat $\Lambda$CDM cosmology with $h=0.73$, 
$\Omega_m=0.27$, $\Omega_{\Lambda}=0.73$.  The redshift of Abell~3827 is 
$cz=29500$~\kms \citep{Struble1999}. The distance is assumed to be 433~Mpc. 
This corresponds to a distance modulus of 38.18~mag and a scale of 
1.74~kpc arcsec$^{-1}$.    
All magnitudes in this paper are in the AB system.  
The mass center of Abell~3827 is assumed to be at RA~$=22^h 01^m 52.90^s$, 
DEC~$=-59^d56^m44.89^s$ \citep{Massey2015}.
We assume the $r$ band solar absolute magnitude to be 4.76~mag \citep{Blanton2003}.  

\section{Data and methods}

We use the datacube obtained by the Multi-Unit Spectroscopic 
Explorer (MUSE) Integral Field Unit (IFU) spectrograph \citep{Bacon2010, 
LeFevre2013} on the European Southern Observatory (ESO) Very Large 
Telescope (VLT). Abell~3827 was observed in 2015 
(ESO programme 295.A-5018(A), PI:  Richard Massey, \citet{Massey2015, Massey2017}), 
centered at RA~$=22^h 03^m 14.65^s$, DEC~$=-59^d56^m43.19^s$.  We reduce and combine 
the data using MUSE Pipeline {\tt muse-2.2} \citep{Weilbacher2014, Weilbacher2016}.  
According to \citet{Massey2017}, observations were taken in dark time with seeing 
around $0.7^{\prime\prime}$.
The total integration time in the final datacube is $3.2$~hr. 
The field of view (FoV) is $1$\amin $\times1$\amin. The wavelength coverage is $475-935$~nm, 
sampled at 1.25\AA/pixel, with mean spectral resolution $\sim 3000$ at the optical 
wavelength range.  The spatial pixel size is $0.2$\asec $\times0.2$\asec.    
Due to the small FoV we limit the sky regions during the 
data reduction, by setting {\tt skymodel\char`_fraction} to $0.01$. 
As a result, the surface brightness profiles in of the MUSE datacube 
are consistent with the \hst/ACS image \footnote{Based on observations made with the 
NASA/ESA Hubble Space Telescope, obtained from the Mikulski Archive at the Space 
Telescope Science Institute, which is operated by the Association of Universities for 
Research in Astronomy, Inc., under NASA contract NAS 5--26555. These observations are 
associated with program 12817.} out to $\mu_{\rm{F606W}}=24$mag~arcsec$^{-2}$, with $\Delta\mu_{\rm{F606W}}\approx0.22$mag~arcsec$^{-2}$ 
at $\mu_{\rm{F606W}}=24$mag~arcsec$^{-2}$.

\begin{figure*}
\vskip 0.15cm
  \centerline{\psfig{file= 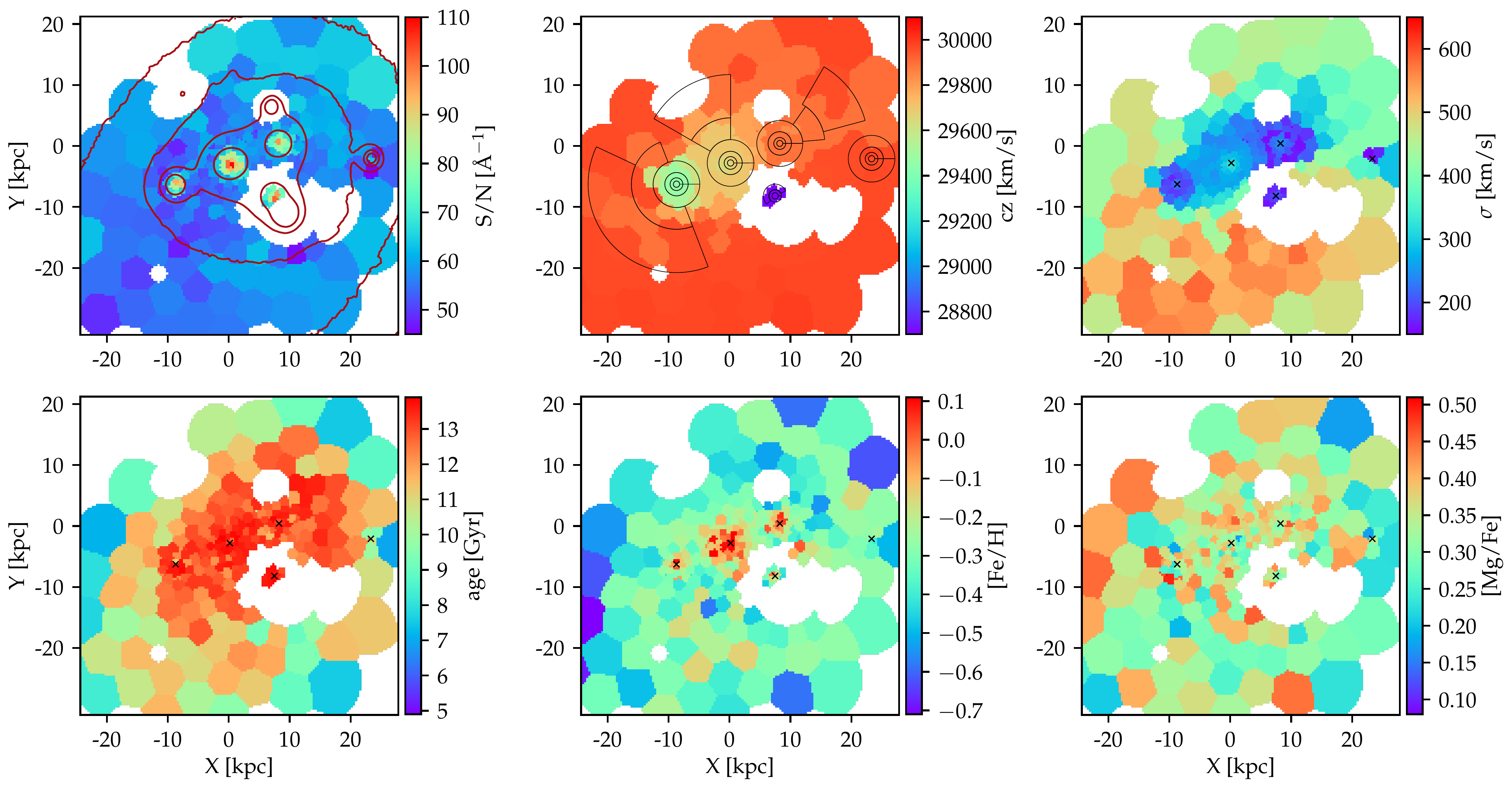, width=18.0cm}}
  \figcaption[A3827map_Summary_May9b.pdf]{
  From top left to bottom right: Spatial distribution of the mean S/N per bin 
  in observed frame $4500$--$5000$~\AA, recession velocity $cz$, velocity 
  dispersions $\sigma_{\star}$, stellar ages, [Fe/H] and [Mg/Fe]. 
  Contours (top left) indicate the surface brightness levels of $\mu_{r}=20$, $21$, $22$ and $23$ \sb~. Black lines (top middle) enclose stacked regions for the radial profiles. Centers of five nuclei galaxies are cross-marked.
  The typical errors of $\log{(\mathrm{age/Gyr})}$, [Fe/H] and [Mg/Fe] are $\approx0.03$~dex in bins with S/N$>100$\AA$^{-1}$.  In regions where $\sigma_{\star}$ is large, the errors are $\approx0.15$~dex in bins with S/N$\sim50$\AA$^{-1}$. 
}
\end{figure*}
\noindent  

Galaxies in Abell~3827 and foreground stars are identified using 
SExtractor \citep{Bertin1996}.  13 ETGs in Abell~3827 are selected for 
further analysis.  They are all confirmed spectroscopically and are 
shown in Figure~1.  We mask out foreground stars and background lensed galaxies 
(whited hatched region).  To study the 
radial profiles of stellar populations, we analyze the spectra by 
binning spaxels in radial directions for N1--N4 and N9.  The binning 
scheme is shown in the right panel of Figure~1. 

To model galaxies spectra we use the absorption line fitter 
\citep[{\tt alf},][]{Conroy2012, Conroy2014, Conroy2018}.  {\tt alf} enables 
stellar population modeling of the full spectrum for stellar ages $>1$Gyr and 
for metallicities from $\sim-2.0$ to $+0.25$.  Parameter space is explored using 
a Markov Chain Monte Carlo algorithm \citep[{\tt emcee},][]{ForemanMackey2013}.  
{\tt alf} adopts the MIST stellar isochrones \citep{Choi2016} and uses a new 
spectral library \citep{Villaume2017} that includes continuous wavelength 
coverage from $0.35-2.4\mu m$ over a wide range in metallicity, which taken 
from new IRTF NIR spectra for stars in the MILES optical spectral library 
\citep{SanchezBlazquez2006}.  Theoretical elemental response functions were 
computed with the ATLAS and SYNTHE programs \citep{Kurucz1970, Kurucz1993}.  
They tabulate the effect on the spectrum of enhancing each of the individual 
elements.  With {\tt alf} in ``full'' mode we fit for parameters including 
a two burst star formation history, the redshift, velocity dispersion, 
overall metallicity, 18 individual element abundances, several IMF parameters 
\citep{Conroy2018}.  Throughout this paper, we use {\tt alf} 
with the IMF fixed to the \citet{Kroupa2001} form.  We use flat priors within 
these ranges: $-10^3$--$10^5$~km/s for recession velocity, $100$--$1000$~km/s 
for velocity dispersion,  $1.0$--$14$~Gyr for age and $-1.8-+0.3$ for 
metallicities.  For each spectrum we fit a continuum in the form of a polynomial 
to the ratio between model and data.  The order of polynomial is 
$(\lambda_{max} - \lambda_{min})/100$\AA.  During each likelihood call the polynomial 
divided input spectrum and model are matched.  The continuum normalization occurs 
in three separate wavelength intervals, $4300-5080$\AA~, $5080-5700$\AA~ and 
$5700-6700$\AA~.  
 
Five galaxies have central velocity dispersion smaller than the resolution of the 
models ($100$~km/s): N5, N6, N11, N12, and N15.  Their spectra are smoothed by 
convolving a wavelength dependent Gaussian kernel with 
$\sigma=\sqrt{100^2-{\sigma_i}^2}$ prior to the modeling, where $\sigma_i$ is 
the wavelength dependent instrumental resolution.  
 
To study the spatial distribution of stellar population parameters, spectra in 
adjacent spaxels are binned by Voronoi tessellation \citep{Cappellari2003}.  
The mean S/N in observed frame $4500$--$5000$\AA~ of this binning scheme is shown 
in the top left panel of Figure~2.  The S/N is between $40$\AA$^{-1}$ and 
$110$\AA$^{-1}$.  In addition to foreground stars and lensed galaxies, 
We also exclude bins where foreground and background 
galaxies overlap with each other, e.g., some spaxels between N4 and N2.  
Only bins where the best-fit spectra and residuals are visually consistent with 
data and data uncertainties are shown here.   
The low S/N at the outskirts ($R>3$~kpc) of low mass ETGs makes it hard to derive reliable 
mass to light ratio (M/L).  Therefore, we assume constant M/L as a function of radius 
and adopt the M/L measured within $R=1$~kpc.  
The stellar masses within $R=5$~kpc of the 13 ETGs are derived by multiplying the 
rest-frame $r$ band total integrated luminosity within $R=5$~kpc, by the best-fit 
M/L in $r$ band within $R=1$~kpc. 

\begin{figure*}
\vskip 0.15cm
\centerline{\psfig{file=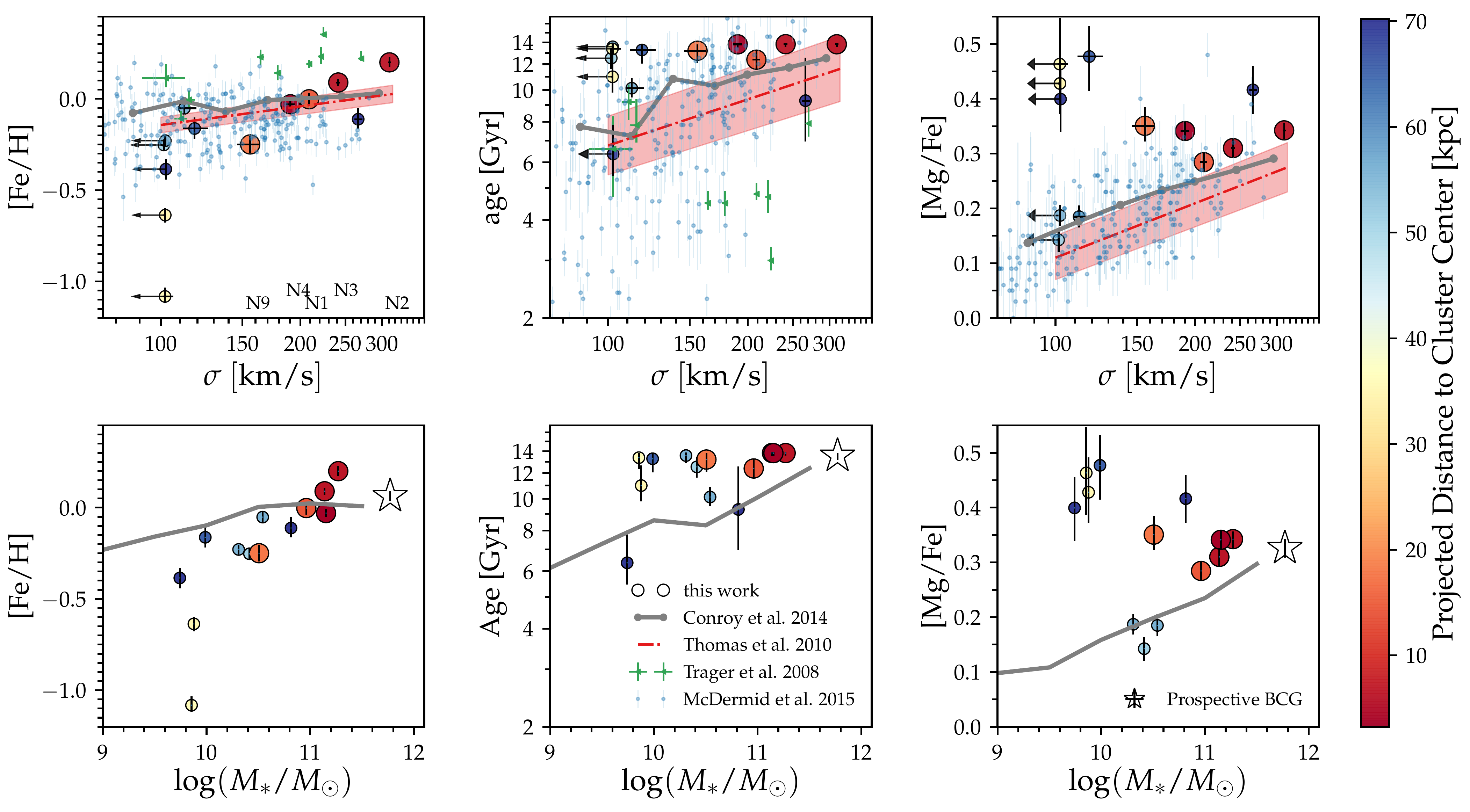, width=18.0cm}}
\figcaption[A3827_sigma_Apr25_10kpc.pdf]{
 Top: Relationships between $\sigma_{\star}$ and stellar population parameters: 
 [Fe/H], stellar ages and [Mg/Fe]. Bottom: Relationships between stellar mass 
 measured within $R=5$~kpc and stellar population parameters measured within 
 $R=1$~kpc.  Colors indicate the projected distance from the center of galaxies 
 to the mass center of the cluster.  13 ETGs in our sample are compared to a 
 large sample of morphologically selected SDSS ETGs in \citet{Thomas2005} 
 (red dash--dotted line), ETGs in ATLAS$^{3D}$ by \citet{McDermid2015} 
 (blue circles), 12 ETGs in the Coma cluster by \citet{Trager2008} 
 (green triangles) and updated results from stacked SDSS ETGs that are binned 
 in $\sigma_{\star}$ ans stellar mass \citep{Conroy2014}. 
}
\end{figure*}
\noindent  

\section{Results and Discussion}
We present results in this section.  Objects N1, N2, N3, N4 and N9 
have the closest projected distances to the mass center ($\leq18$kpc).  Their 
recession velocities relative to the cluster center ($cz=29500$~km/s) 
are $cz=-45^{+4}_{-3}$~km/s, $167^{+3}_{-3}$~km/s, $359^{+3}_{-3}$~km/s, 
$-830^{+4}_{-3}$~km/s and $452^{+4}_{-4}$~km/s, 
respectively.  Given their small projected distances to the cluster center and 
very similar velocities, they are on their way to form a massive 
BCG in the near future.  If we estimate the time it would take based the 
dynamical friction time \citep{Binney1987}, the 5 ETGs will merge within $1$~Gyr.   Therefore we assume N1--N4 and N9 are building 
blocks of a prospective BCG with $\log{(M_{\star}/M_{\odot})}\approx11.7$.   

\begin{figure*}
\vskip 0.15cm
\centerline{\psfig{file=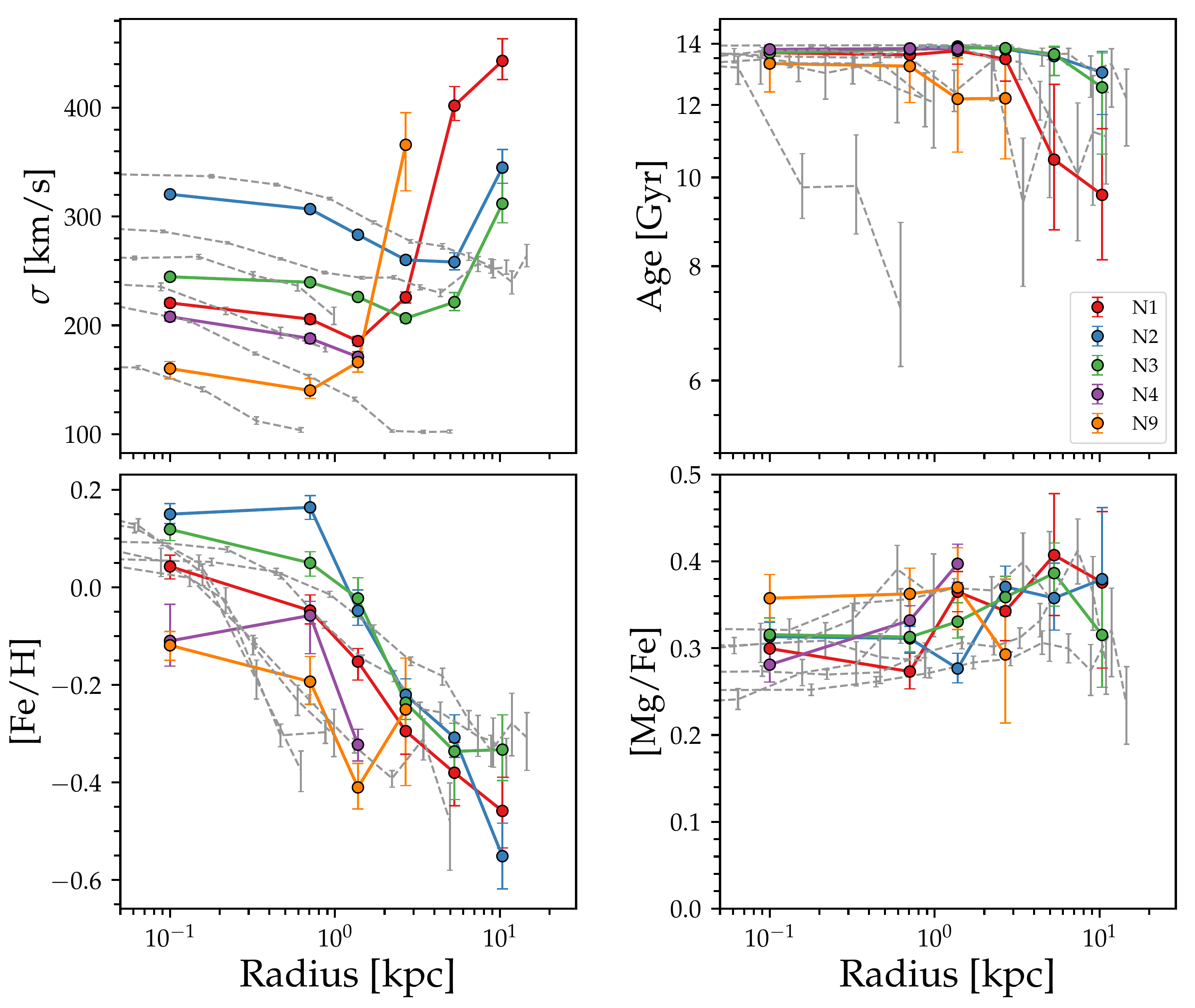, width=15.6cm}}
 \figcaption[A3827_gradient_Apr25b_newmwimf.pdf]{
  Radial profiles of stellar population parameters.  From top left to bottom 
  right: stellar velocity dispersions $\sigma_{\star}$, stellar ages, [Fe/H] 
  and [Mg/Fe] as functions of distance to the center of five galaxies that 
  are closest to the cluster center.  Gray dashed lines represent the radial 
  profiles of six ETGs in \citet{vanDokkum2017b}. 
  }
\end{figure*}
\noindent  

Spatial distribution of the following parameters are shown in Figure~2: 
$cz$, $\sigma_{\star}$, stellar age, [Fe/H] and [Mg/Fe], focusing on 
$D\approx\pm25$~kpc around the cluster center.  The $cz$ distribution shows 
that the five ETGs are all members of the cluster.  The $\sigma_{\star}$ 
map shows that $\sigma_{\star}$ declines outwards from the center of N1--N4 
within $R\sim2$~kpc, and rise towards the outskirts, reaching to 
$\sim400-500$~km/s.  This shows that the stars in the outskirts are tracing 
the gravitational potential of the cluster, instead of any individual 
galaxy, and describes the formation of the cD envelope, or the intracluster 
light component that have been measured in other galaxy clusters 
\citep[e.g.][]{Kelson2002}.  Due to the large $\sigma_{\star}$, the typical 
errors of $\log{(\mathrm{age/Gyr})}$, [Fe/H] and [Mg/Fe] are $\sim0.1$~dex 
in bins with S/N$\approx50$\AA$^{-1}$. The stellar age distribution shows this 
region is uniformly old.  From the [Fe/H] distribution we see declining [Fe/H] 
from the galaxy centers outwards, indicating that the inner regions 
of the galaxies are more metal rich than the cD envelop where the stellar 
content is likely from disrupted low mass systems.  The distribution of 
[Mg/Fe] shows that the $\alpha$--abundance in this region are generally 
high, with no particular pattern of differences between the galaxy centers 
and the cD envelope, indicating that the stellar content have short star formation timescales in general.

Figure~3 shows the main result.  We compare the median 
stellar population parameters within $R=1$~kpc, including [Fe/H], stellar 
ages and [Mg/Fe] as a function of galaxy velocity dispersion $\sigma_{\star}$ 
and galaxy stellar mass.  Colors indicate the projected 
distance to the cluster mass center.  Previous 
studies found scaling relations between stellar population properties and 
the central $\sigma_{\star}$: the stellar components in galaxies 
with higher central $\sigma_{\star}$ are older, more metal rich 
and more $\alpha$--enhanced 
\citep[e.g.][]{Trager2000,Thomas2005,Schiavon2007, Graves2009,Conroy2014}.  
We compare our results to previous studies in Figure~3, including a large 
sample of morphologically selected SDSS ETGs in \citet{Thomas2005} 
(red dash--dotted line), ETGs in ATLAS$^{3D}$ by \citet{McDermid2015} 
(blue circles), 12 ETGs in the Coma cluster by \citet{Trager2008} 
(green triangles).  We estimate [Fe/H] in the above literatures using 
Eq~4 in \citet{Thomas2003} assuming the factor $A=0.94$ \citep{Trager2000}. 
We also compare to stacked SDSS ETGs that are binned in $\sigma_{\star}$ 
and stellar mass in \citet{Conroy2014}, except that they are fit with the 
up-to-date {\tt alf} with upgraded response functions.  

For stellar age and [Fe/H], our results agree with the trends found in 
the general ETG populations: galaxies with higher central stellar velocity dispersion 
or larger stellar mass are older and more metal rich.  For [Mg/Fe] our 
results are distinctly different from ETGs in the field.  The [Mg/Fe] of 
galaxies within $\sim 40$~kpc to the cluster center have high [Mg/Fe], 
even indicating a trend that smaller galaxies are more [Mg/Fe] enhanced. If 
we assume the variation of [Mg/Fe] is due to the differences of star 
formation timescale \citep[e.g.][]{Thomas2005}, Figure~3 illustrates that 
compared to galaxies in all environment, the ETGs with lower $\sigma_{\star}$ or stellar masses within $\sim 40$~kpc in Abell~3827 are 
quenched as early (if not more) as more massive galaxies.  Our results 
highlight the effect of ``environmental quenching'' \citep[e.g.][]{Peng2010} 
on low mass galaxies, possibly due to the complex interplay between 
processes such as ram pressure stripping and strangulation.  
The high [$\alpha$/Fe] and old stellar ages make the 
building blocks very different from low mass galaxies in general, and 
are possibly due to early quenching by the dense environment.  

As described earlier N1--N4 and N9 are very likely building blocks of 
a BCG in the future.  Although multiple nucleus are rarely  
observed, it has been predicted by recent simulations that major 
merger plays an important role on buildup of massive galaxies 
\citep[e.g.][]{RodriguezGomez2016}. We estimate the stellar 
population properties of the prospective BCG through weighting 
the properties of the five ETGs by luminosity.  
The lower limit of its stellar mass is estimated 
using the total stellar mass of N1--N4 and N9 within $R=5$~kpc: 
$\log{(M_{\star}/M_{\odot})}\approx11.7$. The diffuse component is 
not included due to the contamination from the lensed galaxies and the foreground stars. 
The predictions are 
shown as black stars in Figure~3. This prospective BCG would fall on 
all the empirical trends between stellar population parameters and 
stellar mass.    

\begin{figure}[hb] 
\vskip 0.15cm
\centerline{\psfig{file=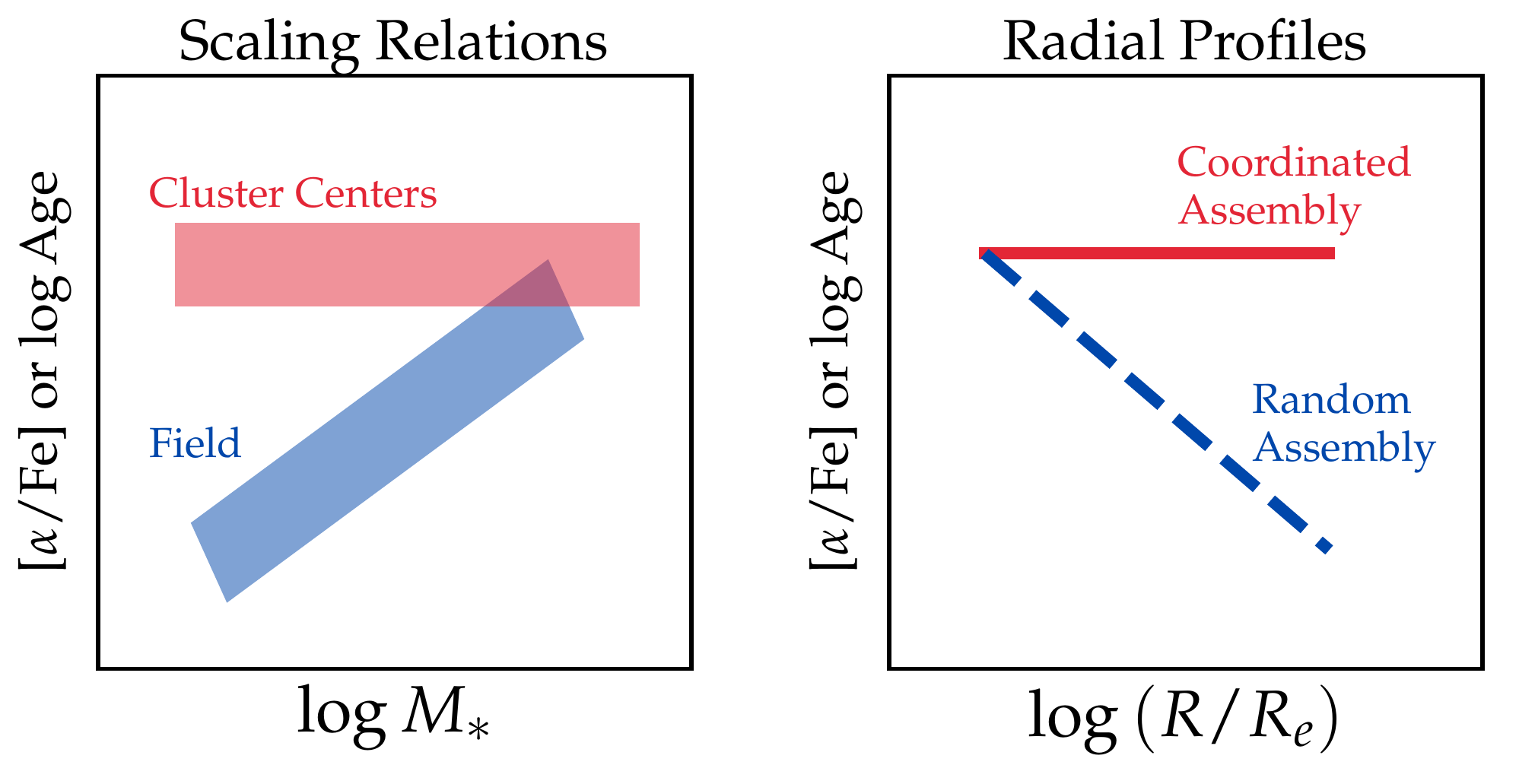, width=8.5cm}}
\figcaption[A3827_cartoon.pdf]{
 Schematic diagrams for scaling relations of [$\alpha$/Fe] or 
 $\mathrm{\log(age)}$ verses stellar mass in different environments (left), 
 and the radial profile of [$\alpha$/Fe] or $\mathrm{\log(age)}$ a massive 
 ETG would build up (right) if it accreted systems randomly in all 
 environments (blue), or coordinately from a sample of low mass galaxies 
 that are quenched early on by the dense environments (red). 
 }
\end{figure}
\noindent  

The coordinated assembly picture can be described by the schematic 
diagrams in Figure~5:  the building blocks of BCGs are low mass 
galaxies in a special environment--cluster centers. 
They follow a relatively flat relation between [$\alpha$/Fe] or stellar age 
and stellar mass (red box), which are distinct from the relations 
followed by galaxies in the fields (blue box).  The right panel 
shows the expected radial profiles of the massive ETGs: As the 
low mass systems in the cluster center accrete onto the outskirts of 
massive ETGs, the coordinated assembly produces flat radial 
profiles of stellar age and [$\alpha$/Fe] (red), whereas if the 
accreted systems are random draws from the low mass galaxy sample 
in all environments the profiles of stellar age and [$\alpha$/Fe] 
would decline with radius.

The radial profiles of stellar population properties confirm the 
coordinated assembly picture.  Figure~4 shows the radial profiles 
of velocity dispersion $\sigma_{\star}$, stellar ages, [Fe/H], and 
[Mg/Fe] for N1--N4 and N9.  The binning schemes are shown in the 
right panel of Figure~1.  These profiles are compared to that of 
6 ETGs in \citet{vanDokkum2017}.  Note that we only show the 
combined profiles from both sides of the 6 ETGs.  Due to the 
contamination of the foreground stars and background lensed galaxies, 
we are not able to extend the radial trends beyond $14$~kpc.  
However, we can already see rising velocity dispersion profiles built up for 
N1--N3, which is consistent with the expected cD galaxy profiles 
\citep[e.g.][]{Kelson2002}.  The stellar ages profiles indicate that the ages of 
the 5 ETGs are generally old from the center to the outskirts.  
[Fe/H] declines with the radius, and seems to depend on the central velocity 
dispersions within $R=2$~kpc.
This is consistent with the two phase formation scenario 
\citep[e.g.][]{Naab2009, vanderWel2014} that the inner regions are mostly 
build up by the dissipational process at high redshifts, thus strongly 
depend on the central stellar velocity dispersions, and the outskirts are 
dominated by accretion of small stellar systems. [Mg/Fe] profiles are 
generally flat. Flat [$\alpha$/Fe] profiles have also been observed 
previously \citep[e.g.][]{Sanchezblazquez2007, Greene2015}. 
This is consistent with the coordinated assembly picture.

\section{Summary}

We have presented detailed stellar population studies for 13 ETGs in 
Abell~3827 using the optical spectra from MUSE on the VLT.  The sample includes five 
ETGs that are involved in an ongoing assembly of a BCG.  Our conclusions 
are summarized as follows:  

\begin{itemize}
\item The 13 ETGs are spectroscopically confirmed 
      members of Abell~3827.  Their stellar age and [Fe/H] fall on empirical 
      trends that galaxies with higher $\sigma_{\star}$ or stellar mass are 
      older and more metal rich.  However, ETGs within $40$~kpc from the 
      cluster center show higher [Mg/Fe] compared to the 
      [Mg/Fe]--$\sigma_{\star}$ and [Mg/Fe]--$M_{\star}$ relations in 
      the field.   
\item From the spatial distribution in the central region of the cluster, 
      the stellar populations in the diffuse stellar light of Abell~3827 
      are generally old and $\alpha$--enhanced.      
\item We show the radial profiles of $\sigma_{\star}$, stellar age, [Fe/H] 
      and [Mg/Fe] that are consistent with previous studies.  The flat 
      stellar age and [Mg/Fe] profiles confirm the coordinated assembly 
      picture.      
\item Our results highlight the effect of ``environmental quenching'', and 
      reveal the coordinated assembly of BCGs: the building blocks of the 
      prospective BCG in Abell~3827 are distinct from the general low mass 
      systems by high [$\alpha$/Fe] due to early quenching by the dense 
      environment.        
\end{itemize}

Future spectroscopic observations of cluster centers will place constraints 
on the formation history of massive galaxies, and will shed additional light 
on the general picture of coordinated assembly.

\acknowledgments
M.G. acknowledges support from the National Science Foundation Graduate 
Research Fellowship.  C.C. acknowledges support from NASA grant 
NNX15AK14G, NSF grant AST-1313280, and the Packard Foundation. 

This research has made use of the services of the ESO Science Archive 
Facility, based on observations collected at the European Organization 
for Astronomical Research in the Southern Hemisphere under ESO 
programme 295.A-5018(A), PI: Richard Massey, and 
based on data obtained from the ESO Science Archive Facility under 
request number 345883.  The computations in this paper were run on the 
Odyssey cluster supported by the FAS Division of Science, Research Computing 
Group at Harvard University.


\begin{thebibliography}{}
\expandafter\ifx\csname natexlab\endcsname\relax\def\natexlab#1{#1}\fi

\bibitem[{Bacon {et~al.}(2010)Bacon, Accardo, Adjali, Anwand, Bauer, Biswas,
  Blaizot, Boudon, Brau-Nogue, Brinchmann, Caillier, Capoani, Carollo, Contini,
  Couderc, Daguis{\'e}, Deiries, Delabre, Dreizler, Dubois, Dupieux, Dupuy,
  Emsellem, Fechner, Fleischmann, Fran{\c c}ois, Gallou, Gharsa, Glindemann,
  Gojak, Guiderdoni, Hansali, Hahn, Jarno, Kelz, Koehler, Kosmalski, Laurent,
  Le~Floch, Lilly, Lizon, Loupias, Manescau, Monstein, Nicklas, Olaya, Pares,
  Pasquini, P{\'e}contal-Rousset, Pell{\'o}, Petit, Popow, Reiss, Remillieux,
  Renault, Roth, Rupprecht, Serre, Schaye, Soucail, Steinmetz, Streicher,
  Stuik, Valentin, Vernet, Weilbacher, Wisotzki, \& Yerle}]{Bacon2010}
Bacon, R., Accardo, M., Adjali, L., {et~al.} 2010, in SPIE Astronomical
  Telescopes + Instrumentation, ed. I.~S. McLean, S.~K. Ramsay, \& H.~Takami
  (SPIE), 773508

\bibitem[{Bertin \& Arnouts(1996)}]{Bertin1996}
Bertin, E., \& Arnouts, S. 1996, Astronomy and Astrophysics Supplement Series,
  117, 393

\bibitem[{Binney \& Tremaine(1987)}]{Binney1987}
Binney, J., \& Tremaine, S. 1987, Princeton, NJ, Princeton University Press,
  1987, 747 p.

\bibitem[{Blanton {et~al.}(2003)Blanton, Hogg, Bahcall, Brinkmann, Britton,
  Connolly, Csabai, Fukugita, Loveday, Meiksin, Munn, Nichol, Okamura, Quinn,
  Schneider, Shimasaku, Strauss, Tegmark, Vogeley, \& Weinberg}]{Blanton2003}
Blanton, M.~R., Hogg, D.~W., Bahcall, N.~A., {et~al.} 2003, The Astrophysical
  Journal Letters, 592, 819

\bibitem[{Cappellari \& Copin(2003)}]{Cappellari2003}
Cappellari, M., \& Copin, Y. 2003, Monthly Notices of the Royal Astronomical
  Society, 342, 345

\bibitem[{Choi {et~al.}(2016)Choi, Dotter, Conroy, Cantiello, Paxton, \&
  Johnson}]{Choi2016}
Choi, J., Dotter, A., Conroy, C., {et~al.} 2016, ApJ, 823, 102

\bibitem[{Conroy {et~al.}(2014)Conroy, Graves, \& van Dokkum}]{Conroy2014}
Conroy, C., Graves, G.~J., \& van Dokkum, P.~G. 2014, The Astrophysical Journal
  Letters, 780, 33

\bibitem[{Conroy \& van Dokkum(2012)}]{Conroy2012}
Conroy, C., \& van Dokkum, P. 2012, The Astrophysical Journal Letters, 747, 69

\bibitem[{Conroy {et~al.}(2018)Conroy, Villaume, van Dokkum, \&
  Lind}]{Conroy2018}
Conroy, C., Villaume, A., van Dokkum, P.~G., \& Lind, K. 2018, ApJ, 854, 139

\bibitem[{Dubinski(1998)}]{Dubinski1998}
Dubinski, J. 1998, The Astrophysical Journal Letters, 502, 141

\bibitem[{Foreman-Mackey {et~al.}(2013)Foreman-Mackey, Hogg, Lang, \&
  Goodman}]{ForemanMackey2013}
Foreman-Mackey, D., Hogg, D.~W., Lang, D., \& Goodman, J. 2013, Publications of
  the Astronomical Society of the Pacific, 125, 306

\bibitem[{Graves {et~al.}(2009)Graves, Faber, \& Schiavon}]{Graves2009}
Graves, G.~J., Faber, S.~M., \& Schiavon, R.~P. 2009, The Astrophysical Journal
  Letters, 693, 486

\bibitem[{Greene {et~al.}(2015)Greene, Janish, Ma, McConnell, Blakeslee,
  Thomas, \& Murphy}]{Greene2015}
Greene, J.~E., Janish, R., Ma, C.-P., {et~al.} 2015, The Astrophysical Journal
  Letters, 807, 11

\bibitem[{Hausman \& Ostriker(1978)}]{Hausman1978}
Hausman, M.~A., \& Ostriker, J.~P. 1978, The Astrophysical Journal Letters,
  224, 320

\bibitem[{Kelson {et~al.}(2002)Kelson, Zabludoff, Williams, Trager, Mulchaey,
  \& Bolte}]{Kelson2002}
Kelson, D.~D., Zabludoff, A.~I., Williams, K.~A., {et~al.} 2002, The
  Astrophysical Journal Letters, 576, 720

\bibitem[{Kroupa(2001)}]{Kroupa2001}
Kroupa, P. 2001, Monthly Notices of the Royal Astronomical Society, 322, 231

\bibitem[{Kurucz(1970)}]{Kurucz1970}
Kurucz, R.~L. 1970, SAO Special Report, 309

\bibitem[{Kurucz(1993)}]{Kurucz1993}
---. 1993, Kurucz CD-ROM, Cambridge, MA: Smithsonian Astrophysical Observatory,
  |c1993, December 4, 1993

\bibitem[{Le~F{\`e}vre {et~al.}(2013)Le~F{\`e}vre, Cassata, Cucciati, Garilli,
  Ilbert, Le~Brun, Maccagni, Moreau, Scodeggio, Tresse, Zamorani, Adami,
  Arnouts, Bardelli, Bolzonella, Bondi, Bongiorno, Bottini, Cappi, Charlot,
  Ciliegi, Contini, de~la Torre, Foucaud, Franzetti, Gavignaud, Guzzo, Iovino,
  Lemaux, L{\'o}pez-Sanjuan, McCracken, Marano, Marinoni, Mazure, Mellier,
  Merighi, Merluzzi, Paltani, Pell{\'o}, Pollo, Pozzetti, Scaramella, Tasca,
  Vergani, Vettolani, Zanichelli, \& Zucca}]{LeFevre2013}
Le~F{\`e}vre, O., Cassata, P., Cucciati, O., {et~al.} 2013, Astronomy \&
  Astrophysics, 559, A14

\bibitem[{Liu {et~al.}(2016)Liu, Peng, Blakeslee, C{\^o}t{\'e}, Ferrarese,
  Jord{\'a}n, Puzia, Toloba, \& Zhang}]{Liu2016}
Liu, Y., Peng, E.~W., Blakeslee, J., {et~al.} 2016, The Astrophysical Journal
  Letters, 818, 179

\bibitem[{Massey {et~al.}(2015)Massey, Williams, Smit, Swinbank, Kitching,
  Harvey, Jauzac, Israel, Clowe, Edge, Hilton, Jullo, Leonard, Liesenborgs,
  Merten, Mohammed, Nagai, Richard, Robertson, Saha, Santana, Stott, \&
  Tittley}]{Massey2015}
Massey, R., Williams, L., Smit, R., {et~al.} 2015, Monthly Notices of the Royal
  Astronomical Society, 449, 3393

\bibitem[{Massey {et~al.}(2017)Massey, Harvey, Liesenborgs, Richard, Stach,
  Swinbank, Taylor, Williams, Clowe, Courbin, Edge, Israel, Jauzac, Joseph,
  Jullo, Kitching, Leonard, Merten, Nagai, Nightingale, Robertson, Romualdez,
  Saha, Smit, Tam, \& Tittley}]{Massey2017}
Massey, R., Harvey, D., Liesenborgs, J., {et~al.} 2017, 1708.04245

\bibitem[{McDermid {et~al.}(2015)McDermid, Alatalo, Blitz, Bournaud, Bureau,
  Cappellari, Crocker, Davies, Davis, de~Zeeuw, Duc, Emsellem, Khochfar,
  Krajnovi, Kuntschner, Morganti, Naab, Oosterloo, Sarzi, Scott, Serra,
  Weijmans, \& Young}]{McDermid2015}
McDermid, R.~M., Alatalo, K., Blitz, L., {et~al.} 2015, Monthly Notices of the
  Royal Astronomical Society, 448, 3484

\bibitem[{Naab {et~al.}(2009)Naab, Johansson, \& Ostriker}]{Naab2009}
Naab, T., Johansson, P.~H., \& Ostriker, J.~P. 2009, ApJ, 699, L178

\bibitem[{Oser {et~al.}(2012)Oser, Naab, Ostriker, \& Johansson}]{Oser2012}
Oser, L., Naab, T., Ostriker, J.~P., \& Johansson, P.~H. 2012, ApJ, 744, 63

\bibitem[{Oser {et~al.}(2010)Oser, Ostriker, Naab, Johansson, \&
  Burkert}]{Oser2010}
Oser, L., Ostriker, J.~P., Naab, T., Johansson, P.~H., \& Burkert, A. 2010,
  ApJ, 725, 2312

\bibitem[{Ostriker \& Tremaine(1975)}]{Ostriker1975}
Ostriker, J.~P., \& Tremaine, S.~D. 1975, The Astrophysical Journal Letters,
  202, L113

\bibitem[{Patel {et~al.}(2013)Patel, van Dokkum, Franx, Quadri, Muzzin,
  Marchesini, Williams, Holden, \& Stefanon}]{Patel2013}
Patel, S.~G., van Dokkum, P.~G., Franx, M., {et~al.} 2013, ApJ, 766, 15

\bibitem[{Peng {et~al.}(2010)Peng, Lilly, Kovac, Bolzonella, Pozzetti, Renzini,
  Zamorani, Ilbert, Knobel, Iovino, Maier, Cucciati, Tasca, Carollo, Silverman,
  Kampczyk, de~Ravel, Sanders, Scoville, Contini, Mainieri, Scodeggio, Kneib,
  Le~F{\`e}vre, Bardelli, Bongiorno, Caputi, Coppa, de~la Torre, Franzetti,
  Garilli, Lamareille, Le~Borgne, Le~Brun, Mignoli, Perez~Montero, Pello,
  Ricciardelli, Tanaka, Tresse, Vergani, Welikala, Zucca, Oesch, Abbas, Barnes,
  Bordoloi, Bottini, Cappi, Cassata, Cimatti, Fumana, Hasinger, Koekemoer,
  Leauthaud, Maccagni, Marinoni, McCracken, Memeo, Meneux, Nair, Porciani,
  Presotto, \& Scaramella}]{Peng2010}
Peng, Y.-j., Lilly, S.~J., Kovac, K., {et~al.} 2010, The Astrophysical Journal
  Letters, 721, 193

\bibitem[{Rodriguez-Gomez {et~al.}(2016)Rodriguez-Gomez, Pillepich, Sales,
  Genel, Vogelsberger, Zhu, Wellons, Nelson, Torrey, Springel, Ma, \&
  Hernquist}]{RodriguezGomez2016}
Rodriguez-Gomez, V., Pillepich, A., Sales, L.~V., {et~al.} 2016, Monthly
  Notices of the Royal Astronomical Society, 458, 2371

\bibitem[{S{\'a}nchez-Bl{\'a}zquez {et~al.}(2007)S{\'a}nchez-Bl{\'a}zquez,
  Forbes, Strader, Brodie, \& Proctor}]{Sanchezblazquez2007}
S{\'a}nchez-Bl{\'a}zquez, P., Forbes, D.~A., Strader, J., Brodie, J., \&
  Proctor, R. 2007, 377, 759

\bibitem[{S{\'a}nchez-Bl{\'a}zquez {et~al.}(2006)S{\'a}nchez-Bl{\'a}zquez,
  Peletier, Jim{\'e}nez-Vicente, Cardiel, Cenarro, Falcon-Barroso, Gorgas,
  Selam, \& Vazdekis}]{SanchezBlazquez2006}
S{\'a}nchez-Bl{\'a}zquez, P., Peletier, R.~F., Jim{\'e}nez-Vicente, J.,
  {et~al.} 2006, Monthly Notices of the Royal Astronomical Society, 371, 703

\bibitem[{Schiavon(2007)}]{Schiavon2007}
Schiavon, R.~P. 2007, ApJS, 171, 146

\bibitem[{Schombert(1988)}]{Schombert1988}
Schombert, J.~M. 1988, ApJ, 328, 475

\bibitem[{Struble \& Rood(1999)}]{Struble1999}
Struble, M.~F., \& Rood, H.~J. 1999, ASTROPHYS J SUPPL S, 125, 35

\bibitem[{Thomas {et~al.}(2002)Thomas, Maraston, \& Bender}]{Thomas2003}
Thomas, D., Maraston, C., \& Bender, R. 2002, arXiv, 897

\bibitem[{Thomas {et~al.}(2005)Thomas, Maraston, Bender, \& Mendes~de
  Oliveira}]{Thomas2005}
Thomas, D., Maraston, C., Bender, R., \& Mendes~de Oliveira, C. 2005, 621, 673

\bibitem[{Thomas {et~al.}(2010)Thomas, Maraston, Schawinski, Sarzi, \&
  Silk}]{Thomas2010}
Thomas, D., Maraston, C., Schawinski, K., Sarzi, M., \& Silk, J. 2010, Monthly
  Notices of the Royal Astronomical Society

\bibitem[{Tinsley(1979)}]{Tinsley1979}
Tinsley, B.~M. 1979, ApJ, 229, 1046

\bibitem[{Tolstoy {et~al.}(2009)Tolstoy, Hill, \& Tosi}]{Tolstoy2009}
Tolstoy, E., Hill, V., \& Tosi, M. 2009, Annu. Rev. Astro. Astrophys., 47, 371

\bibitem[{Trager {et~al.}(2008)Trager, Faber, \& Dressler}]{Trager2008}
Trager, S.~C., Faber, S.~M., \& Dressler, A. 2008, Monthly Notices of the Royal
  Astronomical Society, 386, 715

\bibitem[{Trager {et~al.}(2000)Trager, Faber, Worthey, \&
  Gonz{\'a}lez}]{Trager2000}
Trager, S.~C., Faber, S.~M., Worthey, G., \& Gonz{\'a}lez, J.~J. 2000, The
  Astronomical Journal, 119, 1645

\bibitem[{van~der Wel {et~al.}(2014)van~der Wel, Franx, van Dokkum, Skelton,
  Momcheva, Whitaker, Brammer, Bell, Rix, Wuyts, Ferguson, Holden, Barro,
  Koekemoer, Chang, McGrath, H{\"a}ussler, Dekel, Behroozi, Fumagalli, Leja,
  Lundgren, Maseda, Nelson, Wake, Patel, Labb{\'e}, Faber, Grogin, \&
  Kocevski}]{vanderWel2014}
van~der Wel, A., Franx, M., van Dokkum, P.~G., {et~al.} 2014, ApJ, 788, 28

\bibitem[{van Dokkum {et~al.}(2017{\natexlab{a}})van Dokkum, Conroy, Villaume,
  Brodie, \& Romanowsky}]{vanDokkum2017b}
van Dokkum, P., Conroy, C., Villaume, A., Brodie, J., \& Romanowsky, A.~J.
  2017{\natexlab{a}}, ApJ, 841, 68

\bibitem[{van Dokkum {et~al.}(2017{\natexlab{b}})van Dokkum, Abraham,
  Romanowsky, Brodie, Conroy, Danieli, Lokhorst, Merritt, Mowla, \&
  Zhang}]{vanDokkum2017}
van Dokkum, P., Abraham, R., Romanowsky, A.~J., {et~al.} 2017{\natexlab{b}},
  ApJ, 844, L11

\bibitem[{van Dokkum {et~al.}(2010)van Dokkum, Whitaker, Brammer, Franx, Kriek,
  Labb{\'e}, Marchesini, Quadri, Bezanson, Illingworth, Muzzin, Rudnick, Tal,
  \& Wake}]{vanDokkum2010}
van Dokkum, P.~G., Whitaker, K.~E., Brammer, G., {et~al.} 2010, ApJ, 709, 1018

\bibitem[{Villaume {et~al.}(2017)Villaume, Conroy, Johnson, Rayner, Mann, \&
  van Dokkum}]{Villaume2017}
Villaume, A., Conroy, C., Johnson, B., {et~al.} 2017, ASTROPHYS J SUPPL S, 230,
  23

\bibitem[{Weilbacher {et~al.}(2016)Weilbacher, Streicher, \&
  Palsa}]{Weilbacher2016}
Weilbacher, P.~M., Streicher, O., \& Palsa, R. 2016, Astrophysics Source Code
  Library

\bibitem[{Weilbacher {et~al.}(2014)Weilbacher, Streicher, Urrutia,
  P{\'e}contal-Rousset, Jarno, \& Bacon}]{Weilbacher2014}
Weilbacher, P.~M., Streicher, O., Urrutia, T., {et~al.} 2014, From Stardust to
  Planetesimals, 485, 451

\bibitem[{Worthey \& Collobert(2003)}]{Worthey2003}
Worthey, G., \& Collobert, M. 2003, ApJ, 586, 17

\end{thebibliography}
\end{document}